\documentclass[12pt]{iopart}

\usepackage{comment}
\usepackage[english]{babel} 
\usepackage{microtype} 
\usepackage[svgnames]{xcolor} 
\usepackage{bbold}
\usepackage{bm}
\usepackage{amsmath}
\usepackage{soul}

\usepackage{orcidlink}

\usepackage{hyperref}
\newcommand{\matr}[1]{\ensuremath{\underline{\bm{#1}}\hskip1pt}}

\newcommand{\MF}[1]{\textcolor{black}{#1}}

\newcommand{\rem}[1]{\textcolor{cyan}{\tiny{}}}
\newcommand{\7}[1]{{\bm #1}}

\newcommand{\VW}[1]{\textcolor{black}{#1}}
\newcommand{\rev}[1]{\textcolor{black}{#1}}

\begin{document}
\title[The probability current in the stochastic path integral formalism]{Evaluation of the probability current in the stochastic path integral formalism}

\author{Valentin Wilhelm$^{1}$\footnote{Authors to whom any correspondence should be addressed.}
\, 
Matthias Krüger$^2$\, 
Matthias Fuchs$^1$\, 
Florian Vogel$^{1}\ddag$\, 
}
\address{$^1$ Fachbereich Physik, Universität Konstanz, D-78457 Konstanz, Germany}
\address{$^2$ Institut für Theoretische Physik, Universität Göttingen, 37077 Göttingen, Germany}

\ead{valentin.wilhelm@uni-konstanz.de , florian.vogel@uni-konstanz.de}
\vspace{10pt}
\begin{indented}
\item[]August 2025
\end{indented}

\date{\today}

\begin{abstract}
The  probability current is a vital quantity in the Fokker-Planck description of stochastic processes. It characterizes non-equilibrium stationary states and appears in linear response calculations. We recover and review  the probability current in the Onsager-Machlup functional approach to Markov processes by deriving a self-contained expression  \rev{in general}  non-equilibrium fluctuation-dissipation \rev{relations} using field theoretical methods. The derived formulas \MF{hold for non-constant drift and diffusion tensors and} are explicitly evaluated in an Ornstein-Uhlenbeck process \rev{with non-reciprocal interactions specified as }  a harmonically bound particle in shear flow. \rev{Our work clarifies the concept of the probability current --- familiar from the Fokker-Planck equation --- in the path integral approach. } 

\end{abstract}
\submitto{\jpa}

This is the Accepted Manuscript version of an article accepted for publication in Journal of Physics A: Mathematical and Theoretical. IOP Publishing Ltd is not responsible for any errors or omissions in this version of the manuscript or any version derived from it. This Accepted Manuscript is published under a CC BY licence. The Version of Record is available online at 10.1088/1751-8121/adf534.

\section{\label{sec:level1}Introduction}
The field of classical thermodynamics deals with the laws and principles governing the transformations of many-body systems, such as heat, work, or matter exchange with the environment. These general principles, like the first or second law of thermodynamics, refer to macroscopic quantities. Initially, thermodynamics focused on systems in equilibrium, justified by equilibrium statistical physics. A fundamental idea here is that microstates constituting a macrostate, described by state variables like energy or entropy, are distributed according to Gibbs-Boltzmann statistics. Close to equilibrium, linear response theory allows calculating transport properties through equilibrium correlation functions. At the heart of this is the fluctuation-dissipation theorem (FDT) and Onsager's reciprocal relations \cite{Onsager_Reciprocal_Relations}. In the last three decades, the development of stochastic thermodynamics \cite{ oliveira_2020,Seifert_2012, Solr-871371} and the derivation of new exact relations valid in non-equilibrium systems \cite{Jarzynski_1997}  have extended the validity of thermodynamics deep into the genuine non-equilibrium regime. In thermodynamic equilibrium, the forward and
time-reversed processes occur with the same probability. In contrast, out of equilibrium, one of them is more likely than the other, indicating irreversibility and the production of entropy. Many of these developments are rooted in stochastic processes.

There are three commonly used descriptions of continuous stochastic dynamics. First, the Langevin equation, which provides a microscopic equation of motion for stochastic degrees of freedom. Second, the Fokker-Planck equation determining the evolution of the probability distribution function (pdf). Both are linked via the Kramers-Moyal expansion \cite{KAMPEN2007,Risken}. Lastly, the path integral approach, assigning a probabilistic weight to each trajectory. These three approaches are thought to be equivalent but complementary \cite{Altland_Simons_2010,Kamenev_2011}. For example, the Fokker-Planck approach provides a probability current 
$\7J$, whose vanishing characterizes the equilibrium state. Additionally, in the case of a non-equilibrium steady state (NESS), \MF{the probability current} determines the corrections to the equilibrium fluctuation-dissipation relations \cite{Risken,Agarwal1972}.
The linear response out of equilibrium can not be understood in terms of entropy production alone. The contribution of the \textit{frenetic} components were discussed in \cite{PhysRevLett.103.010602, C4CP04977B}. The correction to the entropic term of the equilibrium FDT has then been identified with the difference of entropic and frenetic effects of the perturbation \cite{Maes_2020}, which hence also constitutes the probability current \cite{Baiesi2009}. 

The richness of the literature and the insights in the origin of the probability current and non-equilibrium response theory make it quite surprising that no derivation of the probability current exists within the framework of the path integral approach, considering that the growing field of stochastic thermodynamics genuinely works at the level of single trajectories.  \rev{In this work, we establish a precise route to define the probability current in the path integral approach.} \VW{To accomplish this, we study the linear response of the system in its path integral formulation and derive a formulation of the general fluctuation-dissipation relation valid for non-equilibrium systems with state dependent diffusivity and of finite dimension.} It turns out to be essential to pay close attention to the discretization of the stochastic integral. We argue that an anti-It\^o discretization leads to the probability current, as it allows expressing the difference of frenetic and entropic components as a state variable of the transition probability. \rem{Entropic and frenetic components had been identified in the path integral response calculation to an applied perturbation,  we derive the probability current in a general linear response theory within the framework of the Onsager-Machlup stochastic path integral. It} 

Deriving this result, we restrict ourselves to Markov processes, where the transition probability does not depend on the history of the system, which is hence said to have \textit{no memory}. Additionally, we assume ergodicity so that the dependence on the initial state also \MF{fades away with time}. In section \ref{sec_FP}, we quickly review the Fokker-Planck description of Markov processes and state the probability current. After introducing the path integral formalism in section \ref{sec_OM}, we present results of the linear response calculation in and out of equilibrium in section \ref{sec_Lin_Rep}. We derive equations known from FP formalism in the path integral formalism and thus re-obtain the expression for the current. Our calculations implicitly exploit that both approaches are based on the same infinitesimal transition probabilities, which are the fundamental objects in both cases. In section \ref{sec_Equivalence}, we build on this fact and show that the correlation functions of general dynamical variables in both theoretical frameworks are the same. Lastly, we consider \rev{the NESS in} a simple system under shear as an example in section \ref{sec_Example}. \rev{Related models of Ornstein-Uhlenbeck processes (OUP) with asymmetric interactions have been formulated for active particles \cite{Szamel2014}, for non-reciprocal interactions \cite{Loos2020},  for  a Brownian gyrator \cite{Abdoli2025}, and in the context of stochastic thermodynamics \cite{Noh2014,Ding2024}.}

\section{Markov-Chains and the Fokker-Planck approach \label{sec_FP}}
A stochastic process is said to possess the \textit{Markov property} if the probability of a future state $\7x_{n+1}$ at time $t_{n+1}$  only depends on the present state. For $t_0<t_1<...<t_{n-1}< t_n$ one can write the conditional probability of the whole process as 
\begin{align}
    P(\7x_{n+1},t_{n+1}|\7x_n,t_n;....;\7x_0, t_0)=  P(\7x_{n+1},t_{n+1}|\7x_n,t_n).
\end{align}
Here, the state $\7x$ is a point in the sample- or phase- space $\mathcal{W}$. Our case of interest is a system of $d$ positional degrees of freedom so that $ \mathcal{W} = \mathbb{R}^{d}$. Instead of labelling the coordinates (e.g.~as $x_\alpha$), we will use (double) scalar products signs $\cdot$ ($:$)  to abbreviate summations, viz.~ $\7a \,\7b : \matr{D} = \sum_{\alpha,\beta=1}^d a_\alpha b_\beta D_{\alpha\beta}$ with a $d\times d$-dimensional matrix $\matr{D}$ ($d$-dimensional vectors are bold). In the following, the index $j\in \{0,1,...,n+1\}$ labels the discretized time and the time ordered sequence $\omega = \{\7x_0,\7x_1,...,\7x_{n+1}\}$ represents a trajectory in the phase space defining a \textit{Markov-chain.} Assuming that \textit{step sizes} in this chain are small and going to the limit of a continuous time variation, one arrives at the Fokker-Planck (FP) equation for the time evolution of the transition probability density \cite{Risken}
\begin{align}
\label{eq_Fokker_Planck_equation}
 \partial_t P(\7x,t|\7x_0,t_0) & =  \big(-\nabla\cdot\bm{\Delta}(\7x,t) +  \nabla \nabla :\;\matr{D}(\7x,t) \big) \;  P(\7x,t|\7x_0,t_0)  \\\notag & \equiv  \Omega(\7x,t) \;  P(\7x,t|\7x_0,t_0)  \;,
\end{align}
with the Kramers-Moyal coefficients, drift vector $\bm{\Delta}$ and diffusion tensor $\matr{D}$, and the Fokker-Planck operator $\Omega$. 
In case of an inhomogeneous  diffusion tensor, it simplifies the notation to introduce a drift  velocity which contains an additional contribution from the gradient of the diffusivity: $\7V(\7x,t) = \bm{\Delta}(\7x,t)-\nabla \cdot  \matr{D}(\7x,t) $. Note that we condition the probability of the process by assuming that the system was in a prepared state at $t_0$ giving the initial value $P(\7x,t_0|\7x_0,t_0)= \bm{\delta}(\7x-\7x_0)$.  The diffusion tensor $\matr{D}(\7x,t)$ is supposed to be positive definite, which inter alia allows to invert it. The Fokker-Planck equation is the starting point of our work. General discussions of the FP equation can be found  in Refs. \cite{KAMPEN2007,Risken}. Later on, starting in Sect.~\ref{sec_Lin_Rep}, we will study the linear response restricting the unperturbed system to be described with time-independent Kramers-Moyal coefficients,  $\bm{\Delta}(\7x)$ and $\matr{D}(\7x)$.  The process is then assumed to be ergodic, i.e, to reach a unique stationary distribution after the transient process.

An example  clarifying the envisioned setting  can be given by a Brownian particle moving in $d$ dimensions. 
The  drift velocity  $\7V$ can be interpreted by a force field $\7F$ via $\7V(\7x)= \frac{1}{k_B T} \,\matr{D}(\7x)\cdot \7F(\7x)$, where we called the prefactor $k_BT $ (thermal energy) to connect to the familiar thermal equilibrium. If the forces are derived from a potential, $\7F(\7x)=- \nabla U(\7x)$, the (ergodic) system relaxes to thermal equilibrium after the transient process, with distribution $p_{\rm eq}(\7x) = \exp{\{-U(\7x)/k_BT}\}/Z$ with the partition sum $Z$. However, our analysis focuses on transient and non-equilibrium steady states, since we include the possibility of non-conservative or non-reciprocal forces. 

The FP equation also takes the form of a continuity equation $\partial_t P =\Omega P = -\nabla\cdot \7J$ in which we defined the probability current 
\begin{align} \label{probcur_def}
    \7J=\bm{\Delta}\; P -\nabla \cdot \matr{D}\;  P= \bm{V} \;P - \matr{D}\cdot \nabla \;  P \;. 
\end{align}
The probability current is an important quantity. Its vanishing in a stationary state is equivalent to satisfied detailed balance and hence to time reversibility, and thus characterizes equilibrium \cite{Risken}; see~\ref{app:FDT} where this connection is worked out. Its vanishing when the potential conditions hold can easily be shown:  
$\7J_{\rm eq}(\7x)= \bm{\Delta}(\7x) p_{\rm eq}(\7x) -\nabla \cdot \matr{D}(\7x)\;  p_{\rm eq}(\7x) = \left( \bm{V}(\7x)  - \frac{1}{k_BT} \matr{D}(\7x) \cdot \7F(\7x) \right)  p_{\rm eq}(\7x) =0
$.

Finally, taking the average of a state observable $A(\7x)$ at time $t$ is given by performing the $d$-dimensional integral over all possible states weighted with the time-dependent pdf 
\begin{align}
\label{Mean_FP}
(A)_{(t)} = \int d\7x A(\7x) P(\7x,t| \7x_0,t_0)\;.
\end{align}
Because of the initial condition of the stochastic process,  the variable takes the value $(A)_{(t_0)}=A(\7x_0)$ at $t_0$. 
Similarly, a correlation function of two observables $A,B$ is given by 
\begin{align}
\label{Korrelation_FP}
(A(t_2),B(t_1))  = \int d \7x_2d \7x_1 A(\7x_2) B(\7x_1) P_2(\7x_2,t_2;\7x_1,t_1|\7x_0,t_0)\;.
\end{align}
The $ P_2(\7x_2,t_2;\7x_1,t_1|\7x_0,t_0)$ is the joint probability distribution of the pairs $(\7x_2,t_2)$ and $(\7x_1,t_1).$ For $t_2>t_1>t_0$ it can be expressed with the solution of the FP equation as 
\begin{align} \label{eq6}
P_2(\7x_2,t_2;\7x_1,t_1|\7x_0,t_0)=P(\7x_2,t_2|\7x_1,t_1) \; P(\7x_1,t_1|\7x_0,t_0)\;.
\end{align}
\rev{Note that averages in the FP approach are indicated by round brackets.} 

\section{Stochastic path integral formalism \label{sec_OM}}

Generally, a path integral formalism can be developed in analogy to the stochastic evolution equation of the pdf, and the approaches are  equivalent to each other \cite{Altland_Simons_2010,Kamenev_2011}. Often, the path integral formalism, which is in many cases the basis in field theory, enables established diagrammatic perturbation theory and allows exploiting numerical tools like the \textit{Path Integral Monte Carlo} method. Thus, it is desirable to recover $\7J$ in this formalism as well. 

There are different methods to express the transition probability for Markov processes as a path integral over all the trajectories leading from the initial $\7x_0$ at $t_0$ to the final state $\7x_{n+1}=\7x$ at $t_{n+1}\equiv t$ \cite{Janssen1976, refId0,PhysRevB.18.353,PhysRevA.8.423,PhysRev.91.1505, Hertz_2017}.
One instructive possibility is to apply the Chapman-Kolmogorov relation for conditional probabilities multiple times 
\begin{align} \label{ChapKol}
    P(\7x_{n+1},t|\7x_0,t_0) =   \int d \7x_{n}  \cdot \cdot \cdot d \7x_{1}   P(\7x_{n+1},t|\7x_{n},t_{n}) \cdot \cdot \cdot P(\7x_1,t_1|\7x_{0},t_{0}).
\end{align}
Here, we set $t_j=j\cdot \Delta t+t_0$.
Taking the limit $t_{i+1}-t_i = \Delta t \to 0$, one finds  
\begin{align}
\label{eq_Transition_Prob_OM}
P(\7x,t|\7x_0,t_0) = \int\limits_{\7x(t_0)= \7x_0}^{\7x(t) = \7x_{n+1}} \mathcal{D} \omega \, \exp{\left(-\mathcal{S}_{\rm OM}\left[\omega\right]\right)}.
\end{align}
Where $\mathcal{S}_{\rm OM}$ is the action, which arises from the discretized Fokker-Planck equation.
It is the so called Onsager-Machlup functional \cite{Risken,Hertz_2017,Wissel1979,Weber_2017_2,Ding2022,dePirey2022}:
\begin{align}
    \mathcal{S}_{\rm OM} = \sum\limits_{j= 1}^{n+1} \bigg\{&\ \frac{\Delta t}{4} \left(\frac{\7x_j-\7x_{j-1}}{\Delta t} -\bm{V}^{(\alpha)}(\overline{\7x}_j,t_j)\right) \cdot  \matr{D}^{-1} (\overline{\7x}_j,t_j)\cdot \left(\frac{\7x_j-\7x_{j-1}}{\Delta t} -\bm{V}^{(\alpha)}(\overline{\7x}_j,t_j)\right) \notag\\ &+ \Delta t \big( \alpha[\nabla_j\cdot\bm{\Delta} (\overline{\7x}_j,t_j)]- \alpha^2 [\nabla_j \nabla_j : \matr{D}(\overline{\7x}_j,t_j)] \big) \bigg\}.
\label{OM}
\end{align}
One integrates over all possible paths from the initial to the final state.
The measure in the above equation reads 
\begin{align}
\label{Definition_Measure}
\mathcal{D}\omega = \underset{n\to\infty}{\text{lim}} \frac{1}{\sqrt{(4\pi \Delta t)^d |\matr{D}(\overline{\7x}_{n+1})|}} \prod_{j = 1}^n\frac{d\7x_j}{\sqrt{(4\pi \Delta t)^d |\matr{D}(\overline{\7x}_j)|}}\;.    
\end{align}
Here, the $|\matr{D}(\overline{\7x}_j)|$ denotes the determinant of the Kramers-Moyal diffusion matrix at state $\overline{\7x}_j$; note that we postulated that it is positive. The limit is understood as $n\to\infty$ and $\Delta t\to0$ such that $t-t_0$ is fixed.  The additional factor to the left of the product sign in Equation \eqref{Definition_Measure} ensures the normalisation $\int d\7x_{n+1} P(\7x_{n+1},t|\7x_0,t_0)=1$.  Different discretization conventions are used.
They enter via 
\begin{align}
    \label{Discretisation_schemes}
\overline{\7x}_j = \alpha \7x_j + (1-\alpha)\7x_{j-1}, 
\end{align} 
for $ \alpha\in \left[0,1\right] $. Derivatives with respect to these mid-points carry an index, viz.~$\nabla_j=\frac{\partial}{\partial \overline{\7x}_j}$, in order to differentiate them from the gradients appearing e.g.~in the FP equation~\eqref{eq_Fokker_Planck_equation}. A dependence on the discretization naturally enters via $\alpha \nabla_j=\frac{\partial}{\partial {\7x}_j}$. It affects the discretized drift velocity, $\bm{V}^{(\alpha)}(\overline{\7x}_j,t_j)=\bm{\Delta}(\overline{\7x}_j,t_j)-2\alpha \nabla_j\cdot \matr{D}(\overline{\7x}_j,t_j)=\bm{V}(\overline{\7x}_j,t_j)+(1-2\alpha) \nabla_j\cdot \matr{D}(\overline{\7x}_j,t_j)$ and the second line in Equation \eqref{OM}. It suggests the useful abbreviation $A^{(\alpha)}(\overline{\7x}_j,t_j)= \alpha\nabla_j\cdot\bm{\Delta} (\overline{\7x}_j,t_j)- \alpha^2 \nabla_j \nabla_j : \matr{D}(\overline{\7x}_j,t_j) $. The second line in Eq.~\eqref{OM} ensures that all different discretizations are equivalent \cite{Ding2022}. 
However, they hold different benefits \cite{Dunkel2009}. From the two most often used ones, the Itô-convention $\alpha=0$ leads to simpler diagrammatics, while the Stratonovich convention $\alpha=\frac{1}{2}$, simplifies considerations involving the time-reversed process. Moreover, the drift velocity takes the value expected from the FP equation for $\alpha=\frac{1}{2}$.  The Hänggi-Klimontovich or anti-Itô discretization  $\alpha=1$ is most useful for connecting to the probability current of the FP approach. The reason for this will become clear in section \ref{sec_Prob_Current}.  

\subsection{On the discretization dependence of the action $\mathcal{S}_{\rm OM}[\omega]$}

 The dependence on the discretization survives in the formal limit of continuous trajectories. For $\Delta t \to 0$, where $\overline{\7x}_j\to\7x(t)$ holds, one can  write the action in Eq.~\eqref{OM} as
\begin{align}
    \notag\mathcal{S}_{\rm OM}[\omega] = \int_{t_0}^t \frac{d \tau}{4} \Bigg\{ \Big[&\dot{\7x}(\tau)- \7V^{(\alpha)}({\7x}(\tau),\tau)+\Big]^T \cdot \matr{D}^{-1}({\7x}(\tau),\tau)  \cdot \Big[ \dot{\7x}(\tau)- \7V^{(\alpha)}({\7x}(\tau),\tau)\Big] \\  & + A^{(\alpha)}({\7x}(\tau),\tau) \Bigg\} \;.
\end{align}
However, this is just a notational abbreviation, making the chosen discretization scheme opaque. It may be worthwhile to illuminate the origin of the discretization dependence starting from the FP equation. Most often it is argued starting from the known ambiguity of the stochastic differential equation \cite{Hertz_2017}. Here, we take a different path and argue that the $\alpha$-ambiguity already enters when deriving the path integral from Eq.~\eqref{ChapKol}. Thus, we consider the short time solution of Eq.~\eqref{eq_Fokker_Planck_equation} for $\Delta t\to0$. It reads: 
\begin{align} \notag
    &P (\7x_j,t_j|\7x_{j-1},t_j -\Delta t) \approx \int  \frac{d \7k}{( 2 \pi)^d} \bigg( 1 + \Delta t \;\Omega(\7x_j,t_j)  \bigg) e^{i\7k \cdot (\7x_j-\7x_{j-1})}\;.
\end{align}
Here, we  used the Fourier-representation of the initial condition, viz.~the Dirac-delta.  
After introducing the shifted positions $\overline{\7x}_j$ according to Eq.~\eqref{Discretisation_schemes}, the  derivatives with respect to $\overline{\7x}_j$ measure the changes of the Kramers-Moyal coefficients during the infinitesimal step. We hence find for the transition probability up to errors in $\Delta t^2$: 
\begin{align}
\label{Eq_Infitesimal_Transition_probability}
     P(\7x_j,t_j|& \7x_{j-1},t_j -\Delta t)  \approx \int \frac{d \7k}{( 2 \pi)^d} \bigg(  1 + \Delta t \Big[ - \alpha [\nabla_j\cdot  \7\Delta(\overline{\7x}_j,t_j)] + \alpha^2[\nabla_j \nabla_j :  \matr{D} (\overline{\7x}_j,t_j) ] \notag\\ 
& - i \7k \cdot \7\Delta (\overline{\7x}_j,t_j)+ 2 i \alpha [ \7k \nabla:\matr{D} (\overline{\7x}_j,t_j)]  - \7k \7k :\matr{D} (\overline{\7x}_j,t_j)
   \Big] \bigg) e^{i\7k \cdot (\7x_j-\7x_{j-1})}\;.
\end{align}
The square brackets around spatial derivatives of the first and second Kramers-Moyal coefficients indicate that the derivative only acts on them.   As $\Delta t \to 0$ is considered, the bracket can be regarded as an expanded exponential  (again up to errors of order $\Delta t^2$)
\begin{align} \label{MSRDJ_inf}
    P(\7x_j,t_j|\7x_{j-1},t_j -\Delta t)  &\approx \int \frac{d \7k}{( 2 \pi)^d}  \;\exp\bigg\{ i \7k \cdot(\7x_j-\7x_{j-1}) -  \\ \notag
& \Delta t \Big[ A^{(\alpha)} (\overline{\7x}_j,t_j)  + i \7k \cdot \7V^{(\alpha)} (\overline{\7x}_j,t_j) + \7k \7k :\matr{D} (\overline{\7x}_j,t_j) \Big]
   \bigg\} \;.
\end{align}
It is written with the abbreviations from above.
The last step to construct the path integral  from the general FP equation with $\alpha$-discretization is to use the Chapman-Kolmogorov Equation \eqref{ChapKol}. 
Performing the calculation directly with Eq.~\eqref{MSRDJ_inf} would lead to a path integral of the \textit{Martin-Siggia-Rose-DeDominici-Jansen} (MSRDJ) formalism \cite{Altland_Simons_2010,Kamenev_2011}. 
To obtain the Onsager-Machlup path integral from Equation \eqref{MSRDJ_inf}, one applies a Hubbard-Stratonovich transformation on the transition probability for an infinitesimal time step (viz.~one performs the Gaussian integral over $\7k$) and uses the Chapman-Kolmogorov procedure  to construct the full path integral. The result is summarized in Eqs.~\eqref{eq_Transition_Prob_OM} to \eqref{Definition_Measure}. The sketch of this derivation justifies its dependence on the discretization. 

As preparation for later,  the derivative of the transition probability with respect to its final position, $\partial/\partial\7 x P(\7x, t|\7x_0,t_0)$ can readily be determined with the present formulae.  As $\7x=\7x_{n+1}$ only enters in the last time-slice, it suffices to consider the short-time transition probability (Eq.~\ref{MSRDJ_inf}) connecting the last two time points:
\begin{align}
   \frac{\partial}{\partial \7x_{n+1}} &
     P(\7x_{n+1}, t_{n+1}|\7x_n,t_n) \approx  \int  \frac{d\7k}{( 2 \pi)^d} \; 
     \bigg[ i \7k  - \Delta t \bigg(  \frac{\partial A^{(\alpha)} (\overline{\7x}_{n+1},t_{n+1}) }{\partial \7x_{n+1}} \\\notag & + i \7k \cdot \frac{\partial \7V^{(\alpha)} (\overline{\7x}_{n+1},t_{n+1}) }{\partial \7x_{n+1}} + \7k \7k : \frac{\partial\matr{D} (\overline{\7x}_{n+1},t_{n+1})}{\partial \7x_{n+1}} \bigg)
   \bigg]   \exp\bigg\{ i \7k \cdot(\7x_{n+1}-\7x_{n})\\ \notag
& + \Delta t \Big[   A^{(\alpha)} (\overline{\7x}_{n+1},t_{n+1})  - i \7k \cdot \7V^{(\alpha)} (\overline{\7x}_{n+1},t_{n+1}) - \7k \7k :\matr{D} (\overline{\7x}_{n+1},t_{n+1}) \Big]
   \bigg\} \;.
\end{align}
The Gaussian $\7k$-integral can be performed by completing the square. In the limit of $\Delta t\to0$, only the factor $i\7k$ survives in the big square bracket in the prefactor of the exponential.  Collecting all terms and building the OM path integral via Eq.~\eqref{ChapKol} leads to the gradient of the pdf
\begin{align} \label{gradient}
      \frac{\partial}{\partial \7x}P(\7x, t|\7x_0,t_0)=   \frac{-1}{2} 
      \int\limits_{\7x(t_0)}^{\7x(t) } \!\!\!
       {\mathcal{D}}\omega & 
    \; 
\Big( \frac{{\7x}_{n+1}-{\7x}_{n}}{\Delta t}
    - \bm{V}^{(\alpha)}(\overline{\7x}_{n+1},t_{n+1}) \Big) \\ & \notag  \cdot \matr{D}^{-1}  (\overline{\7x}_{n+1},t_{n+1})
    \;  e^{-\mathcal{S}_{\rm OM}[\omega]}\;.
\end{align}
This result can be interpreted after multiplying through with the diffusion tensor (possible for $\alpha=0$). It states that a diffusive current (viz.~$\matr{D}\cdot\nabla P$) arises from the expected difference  of the velocity to the current  drift velocity (viz.~$\langle \dot{ \7x}- \7V\rangle$).

\subsection{Averages and correlation functions}

Continuing with the formulation of the path integral approach, averages and correlation functions need to be defined.
Averages over observables in the field theory are marked with pointed brackets and can be calculated according to 
\begin{align}
\langle A(t) \rangle=\int_{\7x(t_0)}
\Tilde{\mathcal{D}}\omega\, A(\7x(t))\exp(-\mathcal{S}_{\rm OM}\left[\omega\right]).
\label{average_OM}
\end{align}
Where $\mathcal{S}_{\rm OM}$ is the action functional of the trajectories $\omega $ leading from the initial configuration $\7x(t_0)$ to the final state $\7x(t)$. Note that  $\Tilde{\mathcal{D}}\omega$ understands also an integration over the final state, in accordance with Equation \eqref{Mean_FP}. Similarly, we define correlation functions for $t_2>t_1>t_0$ as 
\begin{align}
\langle A(t_2), B(t_1) \rangle=\int_{\7x(t_0)}\Tilde{\mathcal{D}}\omega\, A(\7x(t_2)) \; B(\7x(t_1))e^{-\mathcal{S}_{\rm OM}\left[\omega\right]}.
\label{Korrelation_OM}
\end{align}
As in the Fokker-Planck approach, we assume that the system was in a prepared state $ \7x_0=\7x(t_0)$ at $t_0$. 
This is explicitly included in the formalism, since one does not average over the initial state in the equations \eqref{average_OM}, \eqref{Korrelation_OM} and already in \eqref{eq_Transition_Prob_OM}.\\

One should keep in mind that the stochastic path integral is defined for discrete time steps. However, in the following we will also use the continuous limit $\Delta t \to 0$. Nevertheless, this is just a short notation for the discrete process introduced above. Any potentially unclarities can be resolved by going back to the discrete notation.   

\section{Linear Response and probability current\label{sec_Lin_Rep}}

\subsection{Linear Response setting\label{sec_Lin_Rep_setting}}

Now, we have set the stage by introducing the formalism. Hereinafter, we ask how averages are affected by small perturbations applied to the system. In the example of a Brownian particle, we consider a shift in the force field, $\7F \to \7F + \nabla B(\7x)\, h(t)$, by a conservative force modulated by a function of time $h(t)$ that is assumed to render the perturbation small against the unperturbed or reference system. In an equilibrium system, a constant field $h(t)=h^{(s)}$ changes the energy to $U(\7x)- B(\7x) h^{(s)}$.  In the general case, the  time-dependent perturbation gives a change in the Kramers-Moyal drift moment 
\begin{align}\label{stoerung}
\bm{\Delta}(\7x,t)\to\bm{\Delta}(\7x) + \frac{1}{k_BT}\; \matr{D}(\7x) \cdot [\nabla B(\7x)]\;  h(t) \;.
\end{align}
Let us stress again that the prefactor, the inverse thermal energy, is chosen solely in order to allow the interpretation of $B$ and $h$ as set of thermodynamically conjugate variables. Examples would be density and chemical potential, or stress and external strain. In the general case we consider, any other units could be chosen for $B$ and $h$. 
As announced in Sect.~\ref{sec_FP}, the unperturbed system shall, from now on, be characterized by time-independent drift and diffusion terms.  
 We consider a perturbation turned on at $t'\ge t_0$ arbitrarily late after the preparation time $t_0$.  The (negative) change in the action compared to the reference action in first order of $h$ is denoted by $\delta\mathcal{S}_{\rm h}$. In linear response, the perturbation causes the following change in the expectation value of an observable $A$
\begin{align}
\label{eq_general_Expression_Linear_Response}
\langle \Delta A(t)\rangle^{\rm l.r.}&\equiv \langle A(t)\rangle^{\rm l.r.} - \langle A(t) \rangle_0 = \int_{\7x(t_0)}\Tilde{\mathcal{D}}\omega\; A(\7x(t))(e^{\delta\mathcal{S}_{\rm h}[\omega]}-1)e^{-\mathcal{S}_{\rm OM}[\omega]}  
\\\notag&=\int_{\7x(t_0)} \Tilde{\mathcal{D}}\omega\; A(\7x(t))\delta\mathcal{S}_{\rm h}[\omega]e^{-\mathcal{S}_{\rm OM}[\omega]} +\mathcal{O}(h^2)\;. \end{align}
Here, the index $0$ indicates the average over the unperturbed reference system. The lower bound of the integral remains $\7x(t_0)$ even though the perturbation is zero for times smaller $ t'$ since the action $\mathcal{S}_{\rm OM}[\omega]$ still depends on $\7x_0$. 
The change in the action reads 
    \begin{align} \label{deltaS}
    \delta \mathcal{S}_{\rm h} =\sum\limits_{j= 1}^{n+1}  \frac{\Delta t}{2k_BT}h_j \bigg[  \Big(&\frac{\7x_j-\7x_{j-1}}{\Delta t} -\bm{\Delta}(\overline{\7x}_j)\Big) \cdot [\nabla_j B(\overline{\7x}_j)] \\ & \notag - 2 \alpha \; \matr{D}(\overline{\7x}_j): [\nabla_j \nabla_j B(\7x_j)] \bigg] + \mathcal{O}(h^2)\;,
\end{align}
with $h(t_j)=0$ for $t_j<t'$ and abbreviation $h_j=h(t_j)$. Notably, the integrand, i.e.~the perturbed part of the Lagrangian, does not depend on derivatives of $h(t')$ and hence features no potential divergence at $t'$. Thus, we do not have to require that the perturbation is switched on smoothly. Additionally, we remark that the causality of the response function is naturally incorporated in the path integral description. More concretely, due to the normalization of the path integral, one finds for $t_j>t$ 
\begin{align}
	\frac{\partial \langle \Delta A(t)\rangle^{\rm l.r.}}{\partial h_j} =0\;.
\end{align}

Around thermal equilibrium, the linear response is given by the fluctuation-dissipation theorem (FDT) \cite{R_Kubo_1966, MARCONI_2008, Baiesi_2013}
\begin{align}
\label{FDT}
    \langle \Delta A(t)\rangle^{\rm l.r.}_{\rm eq} = 
\frac{1}{k_BT}
    \int_{t'}^{t} h(\tau) \frac{d}{d\tau} \langle A(t) ,B(\tau)\rangle_{\rm eq}\; d\tau \;,
\end{align}
where the bracket on the right-hand side is an equilibrium correlation function, and the FDT relates the magnitude of the perturbation to the time derivative of the correlation function of equilibrium fluctuations;
the prefactor is set by the inverse of the thermal noise strength $k_B T$. 
The FDT is directly related to Onsager's regression hypothesis, stating that the relaxation of a fluctuation caused by a small perturbation driving the system out of equilibrium follows the same laws as the dynamics of fluctuations in the unperturbed thermal system. 

Out of equilibrium, the probability current in the Fokker-Planck description determines the leading correction \cite{Risken, Agarwal1972}  \rev{(note that we use $\langle\rangle$-brackets for the linear response expectations also in the FP framework)}
\begin{align}
\label{FDT_Non_EQ} 
\begin{split}
    \langle\Delta A(t)\rangle^{\rm l.r.}  = \frac{1 }{k_BT}\int_{t'}^{t} d\tau \;h(\tau) \;
    \left( \frac{d}{d\tau} (A(t), B(\tau))_0 - ( A(t), \hat{\7J}(\tau) \cdot [\nabla B(\tau)] )_0 \right)  
\end{split}.\end{align}
with the probability current function $\hat{\7J}$ defined via  $\7J(\7x,t)=\hat{\7J}(\7x,t)P(\7x,t|\7x_0, t_0)$ and ${\7J}$ from Eq.~\eqref{probcur_def}.  Again, the index 0 indicates averaging over the unperturbed probability distribution. Note that the probability flux appears in a measurable quantity in Eq.~\eqref{FDT_Non_EQ} \cite{Seifert_2010, Krueger2009, Kruger2010}. For reference, we provide a derivation of the general fluctuation-dissipation relation (FDR) Eq.~\eqref{FDT_Non_EQ}  in the FP framework in \ref{sec_Appendix_A}. \rev{It was shown in \cite{Chetrite_2008} that this modified FDT close to a stationary non-equilibrium state can be obtained from general fluctuation relations like the Crooks or the Jarzynski relation \cite{Jarzynski_1997, Crooks_1999, Crooks_2000}. Furthermore, the authors of \cite{Chetrite_2008} interpreted  Equation \eqref{FDT_Non_EQ} by showing that the non-equilibrium FDR reduces to the equilibrium FDT when considering the Lagrangian frame moving with $\hat{\7J}$. }
However, practically,  Equation \eqref{FDT_Non_EQ} is often of little use as it requires the knowledge of the probability distribution or the solution of the Fokker-Planck Equation \eqref{eq_Fokker_Planck_equation}. This motivates dynamical approaches  \cite{Maes_2020, Baldovin_2022, Caprini_2021}.  

In the following, we re-derive the probability current in a trajectory-based approach arbitrarily far from equilibrium. For that, we split the perturbation in two terms, $\delta \mathcal{S}_{\rm h}\left[\omega\right] = E\left[\omega\right] + F\left[ \omega\right]$. While this splitting may seem arbitrary at first, it becomes unique when requiring that $E[\omega]$ is the component of $\delta \mathcal{S}_{\rm h}[\omega]$, which can be written as a total time derivative of the field $B$ conjugate to the external perturbation $h$. One can then  write
\begin{align}
\label{Eq_Entropic_Component_Total_Time_derivative}
\int_{\7x_0}^{\7x_{n+1}} &\mathcal{D} \omega \; E[\omega] e^{-\mathcal{S}_{\rm OM}[\omega]}= \frac{1}{2k_BT} \int_{\7x_0}^{\7x_{n+1}} \mathcal{D} \omega  \sum_{j=1}^{n+1}   \Delta t h_j \frac{\Delta B(\overline{\7x}_j)}{\Delta t}e^{-\mathcal{S}_{\rm OM}[\omega]} \; \\ &
\underset{\Delta t \to 0}{\longrightarrow} \frac{1}{2k_BT} \int_{\7x_0}^{\7x_{n+1}}   \mathcal{D}\omega\bigg(h_{t}B(\7x(t)) - h_{t'}B(\7x(t'))-\int_{t'}^{t} d\tau \frac{dh_\tau}{d\tau} B(\7x(\tau)) \bigg)e^{-\mathcal{S}_{\rm OM}[\omega]}.  \notag 
\end{align} 
The first two terms in the last line specify the energy difference between the start and end point, while the integral quantifies the performed work on a trajectory \cite{Baiesi2009}. Hence, $E[\omega]$ can be identified with the excess entropy. Equation \eqref{FDT} suggests that $E\left[\omega\right]$ constitutes the equilibrium FDT stated in Equation \eqref{FDT}. The remainder $F\left[ \omega\right]$ has been labelled the \textit{frenetic component} of the perturbation \cite{PhysRevLett.103.010602,Baiesi_2013}. In the next subsection, we state the discretized path variable expression for $E[\omega]$ and $F[\omega]$ and derive Equation \eqref{Eq_Entropic_Component_Total_Time_derivative}.

\subsection{The manifestation of stochastic calculus in the path integral formalism }

As announced, we decompose the perturbed action $\delta \mathcal{S}_{\rm h}[\omega]$ in the two parts $E[\omega]$ and $F[\omega]$. To be specific, they read for arbitrary  discretizations $\alpha$:
\begin{subequations}
	\label{Entropic_Frenetic_Component}
	\begin{align}
		\label{Eq_27_Entropic_Component}
		E\left[\omega\right]=&\sum\limits_{j=1}^{n+1}    \frac{\Delta t }{2k_B T}h_j \; \bigg\{
		\left( \frac{\7x_j-\7x_{j-1}}{\Delta t} \right) \cdot \nabla_j B(\overline{\7x}_j)-(2 \alpha-1) \matr{D}(\overline{\7x}_j): \nabla_j \nabla_j B(\overline{\7x}_j ) \bigg\}  \;, \\
    \label{Eq_27_Frenetic_Component}
F\left[\omega\right] =&- \sum\limits_{j=1}^{n+1}    \frac{\Delta t }{2k_B T} h_j \; \bigg\{\7\Delta( \overline{\7x}_j) \cdot \nabla_j B(\overline{\7x}_j) + \matr{D}(\overline{\7x}_j): \nabla_j \nabla_j B(\overline{\7x}_j ) \bigg\}. 
	\end{align}
\end{subequations}
The splitting is done such that the discretized expression of $E[\omega]$ leads to Equation \eqref{Eq_Entropic_Component_Total_Time_derivative}, as will be proven in this section. This also substantiates the interpretation that $E[\omega]$  represents the excess entropy due to the perturbation. 

Showing that $E[\omega]$ contains a total time derivative involves proving a generalization of Itô's lemma in the path integral formalism. 
 We prepare our considerations by looking at the total derivative of $B(\7x( \tau))$ with $t_0 \leq  \tau \leq t $:  
\begin{align}
\label{Definition_Total_Derivative}
    \Delta B(\7x(\tau))  =& B(\7x(\tau)) - B(\7x(\tau-\Delta t))   \\ \notag   = &   \Delta \7x \cdot \frac{\partial B(\7x(\tau-\Delta t))}{ \partial \7x (\tau-\Delta t)}  + \frac{1}{2}  \Delta\7x \Delta\7x : \frac{\partial^2B(\7x(\tau-\Delta t))}{\partial \7x(\tau-\Delta t) \partial \7x (\tau-\Delta t)}+ \dots\ .
\end{align}
Here, we wrote $\Delta \7x =\7x(\tau)-\7x(\tau-\Delta t)$. 
The three dots at the end of the previous equation refer to additional terms proportional to $(\Delta \7x)^m$ with $m>  2$. As it turns out, all of these higher-order terms are negligible.  In the standard rules of calculus, one additionally has $\Delta \7x \Delta  \7x\ll \Delta \7x$ for $\Delta t \to 0$. However, this is no longer true in the presence of diffusive motion. To prove this, we look at the general case. Setting $\7x(t)=\7x_j$,  $\7x(t-\Delta t)=\7x_{j-1}$ and $\Delta \7x_j=\7x_j-\7x_{j-1}$, we get from Equation \eqref{Eq_Infitesimal_Transition_probability} for $\Delta t \to 0$:
    \begin{align}
\label{Equation_72_Higher_order_terms}
    \begin{split}
    \int_{\7x_0}^{\7x_j} \mathcal{D}\omega e^{-\mathcal{S}_{\rm OM}[\omega]} (\Delta\7x_j)^m & \overset{\Delta t \to 0}{\longrightarrow}  \int \hspace{-0.17cm} \frac{d \7k}{( 2 \pi)^d} \Tilde{P} \Big[ 1   - \Delta t  \7k \7k: \matr{D}(\overline{\7x}_j) +\Delta t A^{(\alpha)}(\overline{\7x}_j)\Big]  ( \Delta \7x)^m e^{-i \Delta t \7k \cdot \7v_j } \\ 
    & =  -  \int \hspace{-0.17cm}\frac{d \7k}{( 2 \pi)^d} \Tilde{P}  \Delta t  \7k \7k: \matr{D}(\overline{\7x}_j)   ( \Delta \7x_j)^m e^{-i \Delta t \7k \cdot \7v_j } + \mathcal{O}(\Delta t^{m})\\
      & = 2 \Delta t   \int \hspace{-0.17cm} \frac{d \7k}{( 2 \pi)^d} \Tilde{P}   \matr{D}(\overline{\7x}_j)   ( \Delta \7x_j)^{m-2} e^{-i \Delta t \7k \cdot \7v_j } + \mathcal{O}(\Delta t^{m})
      .\end{split}
\end{align}
Here, we have introduced the operator $\Tilde{ P} = \int d \7x_{j-1} P(\7x_{j-1},t_{j-1}|\7x_0,t_0) \big(\cdot\cdot\cdot)$ to shorten the notation. The dots $\big(\cdot \cdot \cdot)$ represent additional terms depending on $\7x_j$ and $\7x_{j-1}$. Additionally, we abbreviated $\7v_j\equiv  -\frac{\7x_j-\7x_{j-1}}{\Delta t}+
\7V^{(\alpha)}(\overline{\7x}_j)$. 
Notably, Eq.~\eqref{Equation_72_Higher_order_terms} holds independent of the discretization. Thus, in the limit $\Delta t \to 0$, the last two equations recover Itô's Lemma under the integral:
\begin{align}
     \int_{\7x_0}^{\7x_j} \mathcal{D} \omega \frac{\Delta B(\7x_j)}{\Delta t} & e^{-\mathcal{S}_{\rm OM}[\omega]} \\ \notag & \overset{\Delta t \to 0 }{ \longrightarrow}  \int_{\7x_0}^{\7x_j}  \mathcal{D} \omega \bigg\{\frac{ \Delta \7x_j}{\Delta t} \cdot  \frac{ \partial B ({\7x}_{j-1})}{\partial \7x_{j-1}}  + \matr{D}({\7x}_{j-1}) :\frac{ \partial ^2B ({\7x}_{j-1})}{\partial \7x_{j-1}\partial \7x_{j-1}} \bigg\} e^{-\mathcal{S}_{\rm OM}[\omega]} 
\end{align}
The evaluation of the terms  at $\7x_{j-1}$ indicates the correspondence  to the Itô-discretization $\alpha=0$  \cite{Baldi_Stochastic_Calculus}. To get the total derivative for an arbitrary discretization $\alpha$, we use $\7x_{j-1}= \overline{\7x}_j-\alpha \Delta \7x_j$. A simple Taylor expansion gives 
\begin{align}
\begin{split}
    \frac{ \partial B ({\7x}_{j-1})}{\partial \7x_{j-1}}&=   \frac{ \partial }{\partial \7x_{j-1}}   B(\overline{\7x}_j- \alpha \Delta \7x_j)  =  \nabla_j B (\overline{\7x}_j)- \alpha  \nabla_j \nabla_j B (\overline{\7x}_j) \cdot  \Delta \7x_j + \mathcal{O}(\Delta t)\;, \\
\frac{\partial^2 B ({\7x}_{j-1})}{\partial \7x_{j-1}\partial \7x_{j-1} }&= \frac{\partial^2 }{\partial \7x_{j-1}\partial \7x_{j-1} } B (\overline{\7x}_j- \alpha \Delta \7x_j)  = \nabla_j \nabla_j B (\overline{\7x}_j)+ \mathcal{O}(\Delta t) \;.
    \end{split}
\end{align}
Here, we anticipated that higher order terms become negligible under the integral as implied by Equation \eqref{Equation_72_Higher_order_terms}. Thus, we find the discretization-dependent generalization of Itô's Lemma known for stochastic differential equations \cite{Hänggi1978}: 
\begin{align} \label{ito}
 &  \int_{\7x_0}^{\7x_j} \mathcal{D} \omega\; \frac{\Delta B(\7x_j)}{\Delta t} e^{-\mathcal{S}_{\rm OM}[\omega]} \overset{\Delta t \to 0 }{\longrightarrow} \\\notag
 & \int_{\7x_0}^{\7x_j} \mathcal{D} \omega \Big\{  \frac{\Delta \7x_j}{\Delta t} \cdot  [\nabla_j B (\overline{\7x}_j)]  + (1-2 \alpha ) \matr{D}(\overline{\7x}_j) : [\nabla_j \nabla_j B  (\overline{\7x}_j)] \Big\} e^{-\mathcal{S}_{\rm OM}[\omega]}  \;. 
\end{align}
Notably, the correction term caused by shifting the arguments vanishes for $\alpha=1/2$. This is why the Stratonovich convention is convenient for studying the entropic $E[\omega]$ and the frenetic $F[\omega]$ components \cite{Seifert_2012, Baiesi2009, dePirey2022}. However, we proceed considering general $\alpha$--discretizations. Equation \eqref{ito} provides the searched-for derivation of Eq.~\eqref{Eq_Entropic_Component_Total_Time_derivative}; viz~it shows that $ \frac{\Delta B(\7x_j)}{\Delta t} \to  \frac{d B(\7x(\tau)}{d \tau}  $ appears in the $E[\omega]$ of Eq.~\eqref{Eq_27_Entropic_Component}, and thus, justifies Eq.~\eqref{Eq_Entropic_Component_Total_Time_derivative}. It also provided the rationale for splitting $\delta \mathcal{S}_{\rm h}$ into $E+F$ in Eq.~\eqref{Entropic_Frenetic_Component}. The general linear response result Eq.~\eqref{eq_general_Expression_Linear_Response} can now be written using Eqs.~\eqref{Eq_27_Entropic_Component} and \eqref{ito} as 
\begin{align}
\label{Equation_Eq_General_FDT}
\langle \Delta A(t)\rangle^{\rm l.r.} =  \int_{t'}^t d\tau \frac{h(\tau)}{k_B T} \frac{d}{d \tau }&\langle
A(\7x(t)), B(\7x(\tau)) \rangle_0 \\& \notag+  \int_{\7x(t_0)}\Tilde{ \mathcal{D} } \omega A(\7x(t)) (F[\omega]-E[\omega])e^{-\mathcal{S}_{\rm OM}[\omega]} .
\end{align}
Here, we identified $(\overline{\7x}_j,t_j) \to \7x(\tau)$. 
In section \ref{sec_Prob_Current}, we show how the second term, including  $(F[\omega]-E[\omega])$, can be expressed with the probability current $\7J$. The case of perturbing around equilibrium can be found in \ref{app:FDT}.

\subsection{Recovering the probability current \label{sec_Prob_Current}}

According to Equation \eqref{Equation_Eq_General_FDT}, the out-of-equilibrium contribution to the FDT of a single trajectory is given by the difference between the entropic and the frenetic component:
\begin{align}
	E[\omega]-F[\omega]= \frac{1}{k_BT} \sum_{j=1}^{n+1}h_j \Delta t X_B^{(\alpha)}(\overline{\7x}_j). \; 
\end{align}
Here, we introduced the $\alpha$-dependent path-variable 
\begin{align}
X_B^{(\alpha)}(\overline{\7x}_j )=& \frac{1}{2}\left( \frac{\7x_j-\7x_{j-1}}{\Delta t} +  \7\Delta(\overline{\7x}_j)  \right)  \cdot [\nabla_j B(\overline{\7x}_j)] +(1-\alpha) \matr{D}(\overline{\7x}_j): [\nabla_j \nabla _j  B(\overline{\7x}_j )] \\\notag 
  =& \frac{1}{2}\left( \frac{\7x_j-\7x_{j-1}}{\Delta t} -  \7V^{(\alpha)}(\overline{\7x}_j) +2  \7V(\overline{\7x}_j)  \right)  \cdot [\nabla_j B(\overline{\7x}_j)]  \\ & \notag +(1-\alpha)\big[\nabla_j \cdot \matr{D}(\overline{\7x}_j)\cdot [\nabla _j  B(\overline{\7x}_j )]\big]
  \;.
\end{align}
Our goal is to express the path variable $X_B^{(\alpha)}[\omega]$ with the state variable $\7J(\7x,t)$ defined in Equation \eqref{probcur_def}. For that, we split up the action as $\mathcal{S}_{\rm OM}[\omega]=\mathcal{S}_{\rm OM}^{(1)}[\omega_{[t_0,t_j]}] +\mathcal{S}_{\rm OM}^{(2)}[\omega_{[t_j,t_{n+1}]}] $ for an arbitrary $t_0 \leq t_j \leq t_{n+1}$; recall the later mapping $t_j\to\tau$ and $t_{n+1}\to t$. We use a similar partitioning for the measure $   \int_{\7x_0}  \Tilde{\mathcal{D}}\omega(\cdot \cdot \cdot)= \int  d \7x_j  \int_{\7x_j} \Tilde{\mathcal{D}}\omega  \int_{\7x_0}^{\7x_j}  {\mathcal{D}}\omega  (\cdot \cdot \cdot)$. 
Equation \eqref{gradient} enables us to replace the discretized velocity in $X_B^{(\alpha)}$ with the gradient of the pdf. Thus, Eq.~\eqref{Equation_Eq_General_FDT} leads to

\begin{align}
	\label{Eq_prob_Current_General_Alpha}
	\notag  &-   \int \Tilde{ \mathcal{D} } \omega A(\7x(t)) (F[\omega]-E[\omega])e^{-\mathcal{S}_{\rm OM}[\omega]}  =   \sum_{j=1}^{n+1} \frac{\Delta t h_j  }{k_BT}  \int_{\7x_0} \Tilde{\mathcal{D}}\omega A(\7x(t)) X_B^{(\alpha)}(\overline{\7x}_j)  e^{-\mathcal{S}_{\rm OM}[\omega]} 
	\\ \notag  =&   \sum_{j=1}^{n+1} \frac{\Delta t}{k_BT}\;  h_j\int d \7x   A(\7x) \int d \7x_j  P(\7x,t_{n+1}|\7x_j, t_j)  \\ &  \notag \times  \bigg\{ (1-\alpha) \int_{\7x_0}^{\7x_j}  \mathcal{D}\omega  \Big[  \nabla_j \cdot \matr{D}(\overline{\7x}_j) \cdot [\nabla_j B(\overline{\7x}_j)] \Big]  e^{-\mathcal{S}^{(1)}_{\rm OM}[\omega_{[t_0,t_j]}]} 
  \\&  + \int_{\7x_0}^{\7x_j}  \mathcal{D}\omega\Big( \7V(\overline{\7x}_j
	)   + \frac{1}{2}\left( \frac{\7x_j-\7x_{j-1}}{\Delta t} -  \7V^{(\alpha)}(\overline{\7x}_j)\right) \Big) e^{-\mathcal{S}^{(1)}_{\rm OM}[\omega_{[t_0,t_j]}]}
  \cdot [\nabla_j B(\overline{\7x}_j)] \bigg\}\,. 
\end{align}
Here, we wrote $\int_{\7x_j}^{\7x} \mathcal{D}\omega \exp\{-\mathcal{S}_{\rm OM}^{(2)} [\omega_{[t_j,t_{n+1}]}]\}=P(\7x,t_{n+1}|\7x_j,t_j)   $.
It is impossible to write the second and third lines concisely with the transition probability $P(\7x_j,t_j|\7x_0,t_0)$ for general $\alpha$. However, this is different in the Hänggi-Klimontovich or anti-Itô discretization  $\alpha=1$. In this convention, the third line vanishes and, using Eq.~\eqref{gradient}, we recover 
\begin{align} \label{FDR}
&-   \int \Tilde{ \mathcal{D} } \omega A(\7x(t)) (F[\omega]-E[\omega])e^{-\mathcal{S}_{\rm OM}[\omega]}  =  \sum_{j=1}^{n+1} \frac{\Delta t h_j}{k_B T}  \int d \7x  \int d \7x_j \, A(\7x) \\ &\times
P (\7x,t_{n+1}|\7x_j,t_j)  \big\{\7\Delta(\7x_j)P(\7x_{j},t_{j}|\7x_0,t_0)  - [\nabla  \cdot \matr{D}(\7x_j) P (\7x_j,t|\7x_0,t_0)]  \big\} \cdot [\nabla B(\7x_j)]  \notag \\ \notag 
&\overset{\Delta t \to 0}{\longrightarrow}\frac{1}{k_BT } \int_{t_0}^t d \tau h(\tau) \langle A(\7x(t)), \hat{\7J}(\7x(\tau)) \cdot [\nabla B(\7x(\tau))] \rangle_0\;.
\end{align}
Here, the $\alpha=1$ convention is indicated by the arguments only depending on $\7x_j$, the later time point in the infinitesimal transition pdf (Eq.~\ref{MSRDJ_inf}).  Notably, the square bracket in the second line indicates that the spatial derivative $\nabla$ only acts on the diffusion tensor and the transition probability, i.e. $ \nabla \cdot \matr{D} P $. 

In the last line, we identified the probability current in the path integral formalism: 
\begin{align} \label{gl36}
	\7J(\7x(\tau)) =&\hat{\7J}(\7x(\tau))P(\7x(\tau), \tau|\7x_0,t_0)\;,
	\\ \notag \hat{\7J}(\7x(\tau)) =&  \bm{\Delta}(\7x(\tau)) - \nabla \cdot \matr{D}(\7x(\tau)) - \matr{D}(\7x(\tau)) \cdot \nabla\log\{P(\7x(\tau) , \tau|\7x_0,t_0)\}.
\end{align}
Here, the \VW{reformulation of Eq.~\eqref{probcur_def} in the} second line \VW{applies where}\rem{ applies only if} $P(\7x , \tau|\7x_0,t_0) > 0$ holds. \VW{When ergodicity is assumed,}\rem{Due to the assumed ergodicity,} this is always true after waiting long enough. It is reassuring that Eq.~\eqref{gl36} agrees with the   Fokker-Planck linear response theory result derived in \ref{sec_Appendix_A}. In \cite{Risken}, the linear response was calculated around a steady state. The assumption of a time-independent probability distribution was essential for the derivation. In our consideration, the transient regime becomes accessible.

We have successfully derived the probability current in the path integral formalism. Setting $\alpha=1$ was essential to recover the probability current $\7J$ as a state variable and its Markovian character. In the next section, we further comment on the relation between the correlation functions in both approaches.

\section{Equivalence of the stationary correlation functions \label{sec_Equivalence}}

The previous sections have implicitly exploited a deeper relation of the Fokker-Planck and the functional integral formalism. As described in Sect.~\ref{sec_OM}, in both cases the transition probability for infinitely small time steps $\Delta t$ reads 
\begin{align}
    P(\7x,t+\Delta t| \7x',t)=&\exp{\left(\Omega(\7x, t)\Delta t+ \mathcal{O}(\Delta t^2) \right)} \delta(\7x-\7x')\;,
\end{align}
where the Dirac distribution $\delta(\7x-\7x')$ enforces that the two states are equal at zero time-difference. This short time expansion combined with the steps described in Sect.~\ref{sec_OM}
leads to 
the functional integral expression \eqref{OM}. 
Hence, the transition probabilities in both formalisms are the same. Secondly, $p_{\rm st}(\7x)= \lim\limits_{t_0\to - \infty} P(\7x, t|\7x_0,t_0)$ is generally true for a fully connected system with time-independent $\Omega(\7x)$. This is the so-called H-theorem, valid for ergodic systems \cite{Chapman}. Hence, $p_{\rm st}(\7x)$ obtained from $ \lim\limits_{t_0\to - \infty} P(\7x, t|\7x_0,t_0)$ is a solution of the Fokker-Planck equation. Thus, it is not surprising that the averages and the correlation functions introduced in the Fokker-Planck and functional integral approaches are  actually fully equivalent. One has
\begin{align}
    (A)_{(t)}=\langle A(t)\rangle\;,\quad \mbox{ and}\qquad
    (A(t_2),B(t_1))=  \langle A(t_2), B(t_1)\rangle\;,
\end{align} 
for e.g.~the correlation functions defined in Eqs.~\eqref{Korrelation_FP} and \eqref{Korrelation_OM}.
This equivalence allows choosing either of the two frameworks depending on the problem at hand. This becomes useful for formal proofs. For example, in the FP approach, the positivity of the spectrum of an autocorrelation function has, to our knowledge, only been proven in equilibrium \cite{Risken}. However, exploiting the path integral formalism allows proving that the spectrum is positive also for stationary autocorrelation functions out of equilibrium. Assuming that the system at the time of the earlier fluctuation, say $t_1$, has already decayed into a stationary state, viz~$P(\7x_1,t_1|\7x_0,t_0) =p_{\rm st}(\7x_1)$, one may take the limit $t_0\to-\infty$ first, and then do a Fourier transformation over a finite time window $2T$. In the limit of $T\to\infty$  one finds (note that here we explicitly assume complex variables and set $B=A^*$)
\begin{align}
    K_{AA}(\nu) &= \lim_{T\to\infty}  \int_{-T}^T  dt_2 dt_1 \langle A(t_2),A^*(t_1) \rangle e^{i\nu(t_2-t_1)}
  \\ & \notag =  \int \Tilde{\mathcal{D}} \omega \left| \int_{-\infty}^\infty dt A(\7x(t)) e^{i \nu t} \right|^2 e^{-\mathcal{S}[\omega]}
    \geq 0\;,
\end{align}
with the frequency $\nu$. This holds since an autocorrelation function in the time domain depends only on the time difference $t_2 - t_1$ due to stationarity.

\section{Example: shear flow \label{sec_Example}}

Our calculations further strengthen the general consensus that the Fokker-Planck approach and the stochastic path integral formalism are equivalent. We continue demonstrating this by an analytical example, the Ornstein-Uhlenbeck process (OUP) of an optically trapped colloid in shear flow \cite{Ziehl2009}. We determine the non-equilibrium corrections to the FDT by calculating the probability current \cite{Kiryl2021}. To be specific, we consider a particle in a harmonic potential with coupling constant $\lambda$. Furthermore, the particle is suspended in a 2D driven viscous fluid. The solvent flows in $\hat{x}$ direction and the strength of the current is proportional to the elongation in the $\hat{y}$ direction. The shear rate $\dot{\gamma} = \frac{\partial u_x}{\partial y}$ of the velocity is assumed constant. This leads to a drift term $\7\Delta  = -\matr{\gamma}\cdot \7r$, linear in the state variable $\7r = (x,y)^T$ and 
\begin{align}
\matr{\gamma} = \begin{pmatrix}
\lambda & -\dot{\gamma}\\ 0 & \lambda
\end{pmatrix}.
\label{drift}
\end{align} 
The superscripted $T$ denotes the transpose of a matrix or vector. Furthermore, we assume a homogeneous diffusion tensor $\matr{D} = D_0 \mathbb{1}_2$ with $\mathbb{1}_2$ the unit $2\times2$ matrix. This model has been studied in the setting of stochastic thermodynamics \cite{Noh2014,Ding2024}.
The probability current has been calculated in the Fokker-Planck approach in \cite{Risken}. There, the density was obtained from a Gaussian Ansatz $p_{\rm st} \propto \exp{(-\7r^T \matr{\sigma}^{-1} \7r/2)}$.  From the FP equation, one finds the variance \begin{align}
    \matr{\sigma}^{-1} = \frac{\lambda}{D_0} \frac{1}{1+\Tilde{\gamma}^2} \begin{pmatrix} 
    1 & -\Tilde{\gamma} \\ -\Tilde{\gamma} & 1+2\Tilde{\gamma}^2 \end{pmatrix}\;,
\end{align}
with $\Tilde{\gamma} = \frac{\dot{\gamma}}{2\lambda}$.
The result of the \rev{stationary}  probability current $\hat{\7 J}_{\rm st}$, which appears in the response function (see Eq.~\ref{FDT_Non_EQ}) then follows as:
\begin{align}
\hat{\7 J}_{\rm st} =\bm{\Delta} - D_0\nabla \log p_{\rm st}  = -\matr{\gamma}\cdot \7r + D_0 \matr{\sigma}^{-1}\cdot \7r.
\label{C}
\end{align}

After recalling the FP textbook results, we turn to the field theory and calculate the transition probability $P(\7r,t|\7r_0,t_0)$ for the path integral approach in the anti-Itô discretization.  \rev{It should be mentioned that the choice of discretization in the example at hand ensures the normalization of the transition probability and thus the calculation of the correct pdf and current. In order to find an $\alpha$-dependent difference between the linear response and the non-equilibrium FDR Eq.~\eqref{Equation_Eq_General_FDT} with \eqref{FDR} would require a nonlinear perturbation $\matr{D}\cdot[\nabla B]$, see Eq.~\eqref{Eq_prob_Current_General_Alpha}.} Nonetheless, in accordance with our considerations in the previous two sections, we choose the anti-Itô convention\rev{, as then the final expression in a linear response calculation generally coincides with the FDR of the FP theory}.
In the Onsager-Machlup path integral formalism, the transition probability reads 
\begin{align}
\begin{split}
    P(\7r,t|\7r_0,t_0)=   e^{2 n \Delta t \lambda}  \int \mathcal{D} \omega  \exp \Bigg\{-\sum_{j=0}^n \frac{1}{4D_0 \Delta t}  \bigg[ \underbrace{\Big(  \mathbb{1}_2 + \Delta t \matr{\gamma} \Big)}_{\equiv \matr{A}} \cdot \;\7r_{j+1}-\7r_j     \bigg]^2   \Bigg\}\;,
    \end{split}
\end{align}
with $\7r_{n+1}\equiv \7r$.
The first factor on the right hand side results from the chosen discretization scheme and is essential for the normalization $\int d\7r  P(\7r,t|\7r_0,t_0) \overset{!}{=}1$. In order to solve the occurring $2n$ integrals, we first perform a Hubbard-Stratonovich transformation and then calculate the transition probability using the MSRDJ path integral 
\begin{align}
\begin{split}
    P(\7r,t|\7r_0,t_0)= &  \int \mathcal{D} \7R  \mathcal{D} \7K \int \frac{d \7k_{n+1}}{4 \pi^2}e^{-\mathcal{S}_{MSRDJ}[\7R,\7K]}e^{2 n \Delta t \lambda}\;, \\
    \mathcal{S}_{MSRDJ}[\7R,\7K]=& \sum_{j=0}^n \left(i
    \7k_{j+1}^T \cdot\Big[ \matr{A} \cdot\7r_{j+1}- \7r_j \Big] + \Delta t D_0 |\7k_{j+1}|^2 \right) \;.
    \end{split}
\end{align}
Here, we defined $\7R^T=(\7r^T_1,...,\7r^T_n)$ and $\7K^T=(\7k_1^T,...,\7k^T_n)$, where the 2D real variable $\7k$ is the so-called response field of the MSRDJ functional. Furthermore, we set 
$\mathcal{D}\7R=\prod_{j=1}^{n} d\7r_j$ and $\mathcal{D}\7K=\prod_{j=1}^{n} \frac{d\7k_j}{(2\pi)^2}$. To perform the first $4n$ integrals, we introduce the $4n\times 4n$ matrix
\begin{align}
   \matr{M}=\begin{pmatrix}
0 \cdot \mathbb{1}_{2n } & i\matr{B}^T\\ i\matr{B}  & 2 D_0 \Delta t \cdot   \mathbb{1}_{2n }
\end{pmatrix}\;,  
\end{align}
where the matrix $\matr{B} \in \mathbb{R}^{2n \times 2n }$ can be written as a $n \times n$ matrix with the matrix $\matr{A}$ on its diagonal entries and the negative identity matrix ${-}\mathbb{1}_{2}$ on the first lower off-diagonal line. The remaining entries are zero. The action can be rewritten as
\begin{align}
\begin{split}
    \mathcal{S}_{MSRDJ}=& \frac{1}{2}(\7R^T,\7K^T)\cdot\matr{M} \cdot\begin{pmatrix}
 \7R \\  \7K
\end{pmatrix}-i \7y^T \cdot \begin{pmatrix}
 \7R \\ \7K
\end{pmatrix} \\&+ i \7k^T_{n+1} \cdot \matr{A} \cdot \7r_{n+1} - \Delta t D_0 |\7k_{n+1}|^2 \;,
\end{split}\\
\7y^T=&(\underbrace{0,...,0}_{n-1\; },\7k_{n+1},\7r_0,\underbrace{0,...,0}_{n-1}).
\end{align}
Due to the simple structures of the matrix $\matr{M}$ and the bi-diagonal matrix $\matr{B}$ one can determine their inverses. The inverse of $\matr{B}$ can be calculated with the Neumann series and the inverse of $\matr{M}$ then follows immediately from the well-known formulae for the inversion of $2\times 2 $ block matrices. After performing the integrals over $\7R$ and $\7K$, one ends up with
\begin{align}
     P(\7r,t|& \7r_0,t_0) = \int \frac{d \7k_{n+1}}{4 \pi^2} e^{-i \7k^T_{n+1} \cdot\matr{A}\cdot\7r_{n+1} -2i \7k_{n+1}^T\cdot \matr{G}(t-t_0)\cdot\7r_0} \\ \notag & \times e^{ - D_0\7k_{n+1}^T \cdot\int_{0}^{t-t_0} dt'\matr{G}(t') \cdot\matr{G}(t')^T\cdot \7k_{n+1}}  e^{\Delta t \lambda} \;.
\end{align}
Here, the Green's function reads in the limit $\Delta t \to 0$
\begin{align}
    \matr{G}(t-t_0) = \matr{A}^{-n  }\underset{\Delta t \to 0}{=} e^{-\lambda (t-t_0) } \begin{pmatrix}
1 & \dot{\gamma}(t-t_0) \\0 & 1 
\end{pmatrix} \;. 
\end{align}
It serves as a check that this expression for the transition probability is equivalent to the one derived in the Fokker-Planck approach \cite{Risken} at least for $\Delta t \to 0$. Since the dependency on the initial state vanishes exponentially, we find for $t_0 \to - \infty $ the correctly normalized stationary probability distribution
\begin{align}
    p_{\rm st}(\7r)= \underset{ t_0 \to - \infty}{\text{lim}} P(\7r,t| \7r_0, t_0) = \frac{\lambda}{2 \pi D_0} \frac{\exp[- \7r^T \matr{\sigma}^{-1} \7r/2 ]}{\sqrt{1 + \Tilde{\gamma}^2}} \;,
\end{align} 
where the variance of the displacements results from 
\begin{align*}
    \matr{\sigma}= \underset{t_0 \to - \infty} {\text{lim}}2 D_0 \int_{0}^{t-t_0} dt'\matr{G}(t') \cdot\matr{G}(t')^T\;.
\end{align*}
 In summary, for $\Delta t\to0$ and $t_0\to-\infty$, the same probability current (see Eq.~(\ref{C})) as in the Fokker-Planck approach is found from Eq.~\eqref{gl36}. It is noteworthy that the general harmonic case, even for a many-body system, can be solved similarly as the calculation is not restricted to two dimensions \cite{Risken}. Perturbation theory then allows calculating any transition probability to arbitrary precision for systems that can be understood as a perturbed Gaussian systems. Hence, using the tools presented in this section, the probability current can be calculated analytically for such systems.

\section{Conclusion}
We have successfully demonstrated that the non-equilibrium fluctuation-dissipation relation can be rigorously recovered from the path integral formalism, supplementing and complementing previous work \cite{PhysRevLett.103.010602, Baiesi2009}. Included in this extension of the FDT is the non-equilibrium probability current, which naturally emerges from field theory in the extension of the response function to non-equilibrium conditions. Its form had been known in the Fokker-Planck approach, and \rev{working out the conceptual steps to derive it} in a path integral representation is our main result. \rev{Our FDR also covers the transient process of relaxation from a prepared initial condition into the stationary state.} It links path integral and Fokker-Planck theory, supporting once more their equivalence. Our calculations, explicitly Equation \eqref{Equation_Eq_General_FDT}, reveal that the probability current is related to the perturbed action of the time reversed trajectories. Due to the assumed ergodicity, our expression for the probability current converges to the known stationary expression after initial transient processes can be neglected. Generally, ergodicity is the main assumption of our approach. \rev{Extending it to driven systems, with time-dependent drift velocity, and to systems with reversible drift currents, seem interesting next steps.} Based on the recent identification of the probability flux in master equations \cite{Baiesi2009}, also the generalization to the corresponding Doi-Peliti functional integral formulation \cite{Altland_Simons_2010} would be of interest. 

\rev{In the path integral approach,} it is vital that the freedom of choosing the discretization most useful for the calculation at hand has to be accompanied by the appropriate correction terms originating from the stochastic nature of the process. Then, the \rev{unique} expression of the Fokker-Planck approach is \rev{recovered}. Interestingly, the Onsager-Machlup path integral is related to the Fokker-Planck equation via the so-called Feynman-Kac formula \cite{Weber_2017_2}. It states that the solution to a second order partial differential equation can be obtained by the expectation value of a solution to a stochastic differential equation, where the latter obeys the rules of stochastic calculus. \rev{Not surprisingly, the path integral calculation also requires a discretization-dependent stochastic chain rule of differentiation (viz.~Eq.~\ref{ito}) known from Itô's lemma. The probability current can then be observed in the Hänggi-Klimontovich discretization ($\alpha=1$).} As an outlook, we note the extension of the functional integral approach to nonlinear response \cite{Caspers2024}, where the role of a finite probability current can now be explored.

\section*{Acknowledgements}
The authors thank Christian Maes for helpful discussions and for comments on the manuscript.
This work is
inspired by discussions during participation in the long-term
workshop "Frontiers in Non-equilibrium Physics 2024" (YITP-T-24-01). MF
would like to acknowledge the warm hospitality during the stay in the
YITP.
The work was supported  by the Deutsche Forschungsgemeinschaft (DFG) via SFB 1432 project C05 and C06.



\appendix
\section{Fokker-Planck linear response with inhomogeneous diffusion \label{sec_Appendix_A}}
For preparing the comparison with the path integral calculation, we provide the derivation of the linear response relation Eq.~\eqref{FDT_Non_EQ} in the FP setting. Starting from the Fokker-Planck equation (Eq.~\ref{eq_Fokker_Planck_equation}) with time-independent Kramers-Moyal coefficients, the perturbation of the drift term given in Eq.~\eqref{stoerung}, leads to a change in the time-evolution operator for $t\ge t'$
\begin{align}
    \delta\Omega(\7x,t) = - \frac{1}{k_BT}\; \nabla \cdot \matr{D}(\7x) \cdot [\nabla B(\7x)]\;  h(t) \;.
\end{align}
With the solution of the unperturbed system: $ P(\7x,t|\7x_0,t_0) =  \exp\{   \Omega(\7x) (t-t_0) \}  \delta(\7x-\7x_0)$, the change in the transition probability linear in the external field $h(t)$ reads
\begin{align}
    \delta P(\7x,t|\7x_0,t_0) = 
    \int_{t'}^{t}  e^{  \Omega(\7x) (t-\tau) }\,   \delta\Omega(\7x,\tau)  \, e^{   \Omega(\7x) (\tau-t') } \,P(\7x,t'|\7x_0,t_0)\; d\tau + {\cal O}(h^2)\;.
\end{align}
The linear change in an arbitrary variable A, defined in Eq.~\eqref{eq_general_Expression_Linear_Response}, thus becomes 
\begin{align}
    \langle\Delta A(t)\rangle^{\rm l.r.}  = \frac{-1}{k_BT}\int_{t'}^{t} d\tau h(\tau)
    \left(  A  e^{  \Omega (t-\tau) }\,\nabla \cdot \matr{D} \cdot [\nabla B]    \, e^{   \Omega (\tau-t') } \right)_{(t')}\, ,
\end{align}
where operators act to anything on the right, including the pdf $P(\7x,t'|\7x_0,t_0)$ that provides the averaging,  if not marked differently by square bracket. A side calculation establishes the required decomposition of the quantity containing $B$ in the last equation. Based on 
\begin{align}
   \left(  B \Omega - \Omega B \right) P &= [\nabla B]   \cdot  \left(  (\bm{\Delta}-\nabla \cdot \matr{D} )  P \right) - \nabla \cdot \left( \matr{D} \cdot [\nabla B]    P \right) \\\notag
   &= [\nabla B]   \cdot  \7J  - \nabla \cdot \left( \matr{D} \cdot [\nabla B]    P \right) \; ,
\end{align}
where the probability current from Eq.~\eqref{probcur_def} appeared, 
one immediately finds with the expression $ \hat{\7 J}(\7x,t') =(1/ P(\7x,t'|\7x_0,t_0)) (\bm{\Delta}(\7x)-\nabla \cdot \matr{D}(\7x) )  P(\7x,t'|\7x_0,t_0) $:
\begin{align}
\label{Equation_A_5}
    \langle\Delta A(t)\rangle^{\rm l.r.}  = \frac{-1}{k_BT}\int_{t'}^{t} d\tau\;  h(\tau) \;  \big( A  e^{  \Omega (t-\tau) } \, \big\{   \Omega B - B \Omega  + [\nabla B]  \cdot \hat{\7 J} \big\} \, e^{   \Omega (\tau-t') } \big)_{(t')} \;.
\end{align}
Now, the pdf that provides the averaging can be propagated up to $\tau$: 
\begin{align}
    P(\7x,\tau|\7x_0,t_0)=e^{   \Omega (\tau-t') }P(\7x,t'|\7x_0,t_0)
\end{align} Now, the first two terms in the curly bracket in Equation \eqref{Equation_A_5} can be identified as the negative $\tau$-derivative. These steps lead to the FDR in the main text, viz.~Eq.~\eqref{FDT_Non_EQ}. 

\section{Equilibrium: The fluctuation-dissipation theorem} \label{app:FDT}

In this appendix,  to keep this manuscript self contained, we reproduce the properties of $E[\omega]$ and $F[\omega]$ under path reversal \cite{C4CP04977B,MAES20201,PhysRevE.88.052109},  from which the FDR is obtained when perturbing around equilibrium, i.e.,  when the unperturbed state obeys detailed balance. $E[\omega]$ is anti-symmetric under reversing the path. 
This is directly apparent  in Equation~\eqref{Eq_27_Entropic_Component} using the Stratonovich convention $\alpha =1/2$, but holds for any $\alpha$. $F[\omega]$ is symmetric under such reversal \cite{C4CP04977B, MAES20201, PhysRevE.88.052109}.  To make this explicit, we introduce 
\begin{align}
     \Theta:  \7x_j \longrightarrow \7x_{n-j} \; , \mbox{ for } \; j={0, \dots, n} \;,
\end{align}
the time reversal operator, mapping the path $\omega$ to the time-reversed one. Kinematic reversal is omitted because we restrict to path symmetric degrees of freedom and observables. 
$E$ and $F$ in Equation \eqref{Entropic_Frenetic_Component} can then be expressed as  \cite{PhysRevLett.103.010602}  
\begin{align}
E\left[ \omega\right] &= \frac{1}{2}\left(\delta \mathcal{S}_{\rm h}\left[\omega\right] -\delta \mathcal{S}_{\rm h}\left[\Theta \omega\right]\right)\\
F\left[ \omega\right] &= \frac{1}{2}\left(\delta \mathcal{S}_{\rm h}\left[\omega\right] + \delta \mathcal{S}_{\rm h}\left[\Theta \omega\right]\right).
\end{align} 
To derive the FDT, we consider here a system which obeys detailed balance, i.e., which, at time $t_0$ is prepared in an equilibrium state, and then, if unperturbed, follows equilibrium dynamics up to time $t$. The equilibrium path weight ${\cal P}(\omega)$, which contains the distribution at time $t_0$ as well as the weight between $t_0$ and $t$, is path reversal symmetric, i.e.,  ${\cal P}(\omega)={\cal P}(\Theta\omega)$.
For simplicity, we set here $t'=t_0$, and obtain for the linear response around equilibrium, 
\begin{align}\label{eq22}
\begin{split}
\langle \Delta A(t)\rangle^{\rm l.r.}_{} &=
\langle A(t) \rangle-\langle A (t)\rangle_{0}
=\langle A(t) \rangle-\langle A(t') \rangle_{}
\\
& =   \int \ \Tilde{\Tilde{\mathcal{D}}}\omega A(\7x(t))( E[\omega]+F[\omega]) {\cal P}(\omega)\\& \,\,\,\,\,- \int  \Tilde{\Tilde{\mathcal{D}}} ( \omega) A(\7x(t))( E[\Theta \omega]+F[\Theta \omega]) {\cal P}(\Theta\omega) \\ &=
 \langle A(t) E(\omega)\rangle_0.
\end{split}
\end{align}
We introduced $\Tilde{\Tilde{\cal D}}$ to indicate averaging over initial and final states at $t'$ and $t$, respectively.
In the third line, we used that  the average at $t'$ equals the average at $t$ when sampling reversed paths. In the last step, we used the mentioned properties, ${\cal P}(\Theta\omega)={\cal P}(\omega)$, $E(\Theta\omega)=-E(\omega)$, and $F(\Theta\omega)=F(\omega)$. Equation~\ref{eq22} is the FDT (i.e.~Eq.~\ref{FDT}), and it  shows that the equilibrium response of a state observable $A$ may be expressed in terms of the entropic component \cite{MAES20201}. One may as well show, using $ E[\omega]-F[\omega]=\delta \mathcal{S}_{\rm h}\left[\Theta \omega\right]$
\begin{align}
     &  \int \ \Tilde{\Tilde{\mathcal{D}}}\omega A(\7x(t))( E[\omega]-F[\omega]) {\cal P}(\omega)= \int \ \Tilde{\Tilde{\mathcal{D}}}\omega A(\7x(t)) \delta \mathcal{S}_{\rm h}\left[\Theta \omega\right] {\cal P}(\omega)\notag\\
     &= \int \ \Tilde{\Tilde{\mathcal{D}}}\omega A(\7x(t')) \delta \mathcal{S}_{\rm h}\left[\omega\right] {\cal P}(\omega)\notag=\langle \Delta A(t')\rangle^{\rm l.r.}_{} =0.
\end{align}
In the last step, we noted that the linear response at $t'$ vanishes. We thus see that the last term in \eqref{Equation_Eq_General_FDT} vanishes, if the unperturbed system obeys detailed balance.

\section*{References}

\bibliography{Literature.bib}

\end{document}